\documentclass[conference]{IEEEtran}
\IEEEoverridecommandlockouts
\usepackage{cite}
\usepackage{amsmath,amssymb,amsfonts}
\usepackage{algorithmic}
\usepackage{graphicx}
\usepackage{textcomp}
\usepackage{subcaption}
\usepackage{multirow}
\usepackage{graphicx}
\usepackage{xcolor}
\def\BibTeX{{\rm B\kern-.05em{\sc i\kern-.025em b}\kern-.08em
    T\kern-.1667em\lower.7ex\hbox{E}\kern-.125emX}}
\begin{document}

\title{Unsupervised Representations Improve Supervised Learning in Speech Emotion Recognition}

\author{\IEEEauthorblockN{Niloufar Faridani$^\dag$}
\IEEEauthorblockA{\textit{Electrical Engineering Faculty} \\
\textit{University of Tehran}\\
Tehran, Iran\\
niloufaridani@ut.ac.ir}
\and
\IEEEauthorblockN{Amirali Soltani Tehrani$^\dag$}
	\IEEEauthorblockA{\textit{Electrical  Engineering Faculty} \\
		\textit{University of Tehran}\\
		Tehran, Iran\\
		soli.tehrani@ut.ac.ir}
\and
\IEEEauthorblockN{Ramin Toosi}
\IEEEauthorblockA{\textit{Electrical  Engineering Faculty} \\
	\textit{University of Tehran}\\
	Tehran, Iran\\
	r.toosi@ut.ac.ir}
}

\maketitle
\def\thefootnote{\dag}\footnotetext{These authors contributed equally to this work}\def\thefootnote{\arabic{footnote}}

\begin{abstract}
Speech Emotion Recognition (SER) plays a pivotal role in enhancing human-computer interaction by enabling a deeper understanding of emotional states across a wide range of applications, contributing to more empathetic and effective communication. This study proposes an innovative approach that integrates self-supervised feature extraction with supervised classification for emotion recognition from small audio segments. In the preprocessing step, to eliminate the need of crafting audio features, we employed a self-supervised feature extractor, based on the Wav2Vec model, to capture acoustic features from audio data. Then, the output feature-maps of the preprocessing step are fed to a custom designed Convolutional Neural Network (CNN)-based model to perform emotion classification. Utilizing the ShEMO dataset as our testing ground, the proposed method surpasses two baseline methods, i.e. support vector machine classifier and transfer learning of a pretrained CNN. comparing the propose method to the state-of-the-art methods in SER task indicates the superiority of the proposed method. Our findings underscore the pivotal role of deep unsupervised feature learning in elevating the landscape of SER, offering enhanced emotional comprehension in the realm of human-computer interactions.
\end{abstract}


\section{Introduction}
Communication establishes the foundation for strong connections between individuals. A community where people feel heard, appreciated, and supported is highly influenced by thoughtful speaking and listening, as well as by taking into account how our words affect others. Speech is one of the most important means through which people communicate with their surroundings. Speaking remains the most natural, prevalent, and effective method of human engagement, despite the fact that modern communication increasingly takes place via keyboards and displays \cite{b1}. When we communicate, using the right emotions may affect how people hear and understand what we are saying while also demonstrating our interest, commitment, and sincerity. This makes our genuine intentions clear and makes it possible for receivers to react in a suitable way. Designing technologies for human usage seems to benefit from using emotion detection \cite{b34} to improve interaction given that human-computer interaction is being used in a variety of applications \cite{b35}. Therefore, a technique and tool for extracting and identifying emotions in speech would be useful.
Speech emotion recognition (SER) seeks to automatically identify an individual's emotional or physical state from their voice. Despite not altering linguistic content, emotional state has a big impact on communication since it gives feedback in different situations \cite{b2}. The growing number of fields that have profited from SER research, including as contact centers, video games, and lie detection, has increased interest in this area \cite{b3}. SER can be used in interactive online learning as well. Computer systems can adaptively choose the most effective methods for presenting the remaining course material by detecting a student's emotional state \cite{b8}. In therapeutic settings, SER enables professionals to better understand patients' mental states and perhaps uncover hidden emotions \cite{b9}. Virtual voice assistants like Siri, Alexa, Cortana, and Google Assistant have proliferated in popularity as human interaction interfaces in recent years \cite{b19}. Systems run the danger of being viewed as cold, socially awkward, unreliable, and incompetent if they are unable to recognize or effectively respond to a user's emotional condition \cite{b20}. SER is therefore becoming a more crucial capability \cite{b19}. To represent emotional cues in voice signals, the SER approach primarily uses feature extraction, which is followed by feature classification to categorize emotions \cite{b4}.
The first stage of SER is feature extraction, which entails finding emotional cue representations in speech signals. A major problem in SER continues to be identifying the best features to achieve improved performance \cite{b15}. Certain aspects must successfully communicate essential emotional information. Source-based excitation features, prosodic features, vocal tract factors, and hybrid feature sets are just a few of the feature types that researchers have developed for speech processing \cite{b5}.Speech contains linguistic and auditory information that conveys emotion. Previous research produced embeddings for challenges involving spoken language using methods such as phone representations \cite{b21} and Mel-frequency approaches \cite{b22}. Energy-related features, pitch frequency features, formant frequencies\cite{b26, b27}, zero crossing rate (ZCR), linear prediction coefficients (LPC), linear prediction cepstral features, mel-frequency cepstrum coefficients (MFCCs) and its first derivative \cite{b28}, RASTA-PLP features \cite{b29}, and others have all been studied to improve emotion classification.Pitch, intensity, and spectral characteristics were used as audio features in \cite{b18} to extract emotion data. For deep learning challenges, new feature learning methods have recently emerged. Given the lack of labeled data for the majority of the 7,000 languages spoken worldwide, many modern language models use self-supervised rather than supervised learning. For the most part, there aren't many hours worth of annotated voice data. As a result, self-supervised approaches facilitate cross-lingual transfer and language learning for low-resource languages \cite{b14, b16}.
In order to improve voice recognition, unsupervised pre-training techniques like Wav2Vec have been developed \cite{b14}. The 1,000 hours that are typically found in supervised academic datasets have been significantly surpassed by these strategies, which have scaled quickly to 1,000,000 hours of training data \cite{b17}. Because models automatically learn from unlabeled voices, this extensive scale is possible. Multilingual transcription systems and other speech communities have long been interested in cross-lingual learning, which tries to create models that use data from different languages to improve performance \cite{b12, b13}.unsupervised pre-training methods have advanced state-of-the-art performance when fine-tuned on benchmark tasks, especially in low-resource environments \cite{b14}.

Feature classification using both linear and nonlinear classifiers is applied in the second stage of SER systems \cite{b6}. Significant interest in voice recognition has been generated by a proposal to improve discrete language models using neural networks \cite{b7}. Speech emotion identification has made revolutionary advances in machine learning recently because of Deep Neural Networks (DNNs). Convolutional Neural Networks (CNNs) use frame-level features to identify patterns and local context. CNNs enable integrated feature representation and classification to be trained end-to-end in one optimization process.
CNNs and long short-term memory (LSTM) networks are two common DNN architectures that recently emerged for voice emotion recognition \cite{b23}. A three-layer LSTM + DNN model for emotional computing was proposed in an early, significant study and trained on the functionals of LLDs \cite{b24}. Stuhlsatz et al. \cite{b25} used generalized discriminant analysis (GerDA) and restricted Boltzmann machines (RBMs) to directly extract discriminative features from raw data.
in \cite{b10}, After a deep neural network feature extractor, a support vector machine (SVM) was used as the final classifier for emotion recognition. Across multiple acoustic feature groups, this architecture intended to learn relevant representations. The proposed model's assertion 64\% accuracy demonstrated better recognition performance.
In contrast to conventional methods that rely on speech segmentation, the fully CNN with attention mechanisms presented in \cite{b30} can process variable-length speech directly. The accuracy of this design in recognizing emotions was 70.4\%. To overcome restrictions caused by insufficient training data, the authors investigated transfer learning. Speech analysis should use careful attention mechanisms since silent speech contains prosodic clues and tonal dynamics.
On the ShEMO dataset, a CNN + BLSTM architecture acting on low-level acoustic descriptors extracted from voice frames obtained 78.29\% accuracy \cite{b11}.

In \cite{b19}, transfer learning for SER is investigated by using representations from a pre-trained wav2vec 2.0 model and basic neural networks. To combine outputs across layers of the pre-trained model, the authors provide trainable weights that are learned along with the downstream model. Using the IEMOCAP and RAVDESS emotion datasets, the performance of wav2vec 2.0 variations with and without speech recognition fine-tuning is examined

In this study, we extract acoustic features from audio data using a wav2vec feature extractor.  Then, a novel audio classification model inspired by CNN is fed these features. The ShEMO emotion recognition dataset serves as the training and evaluating on the ShEMO On this dataset, our proposed method performs well while requiring little to no pre-processing of the raw audio.The adaption of CNN architectures for feature learning from auditory inputs is a significant contribution of our work. Our results show that this CNN-based model outperforms baseline approaches and is successful at extracting emotional cues from speech. In addition, our model offers a generalizable method for speech emotion recognition by relying on deep feature learning as opposed to hand-crafted audio features.

\section{Proposed Method}
\begin{figure*}[htbp]
	\includegraphics[width=\linewidth]{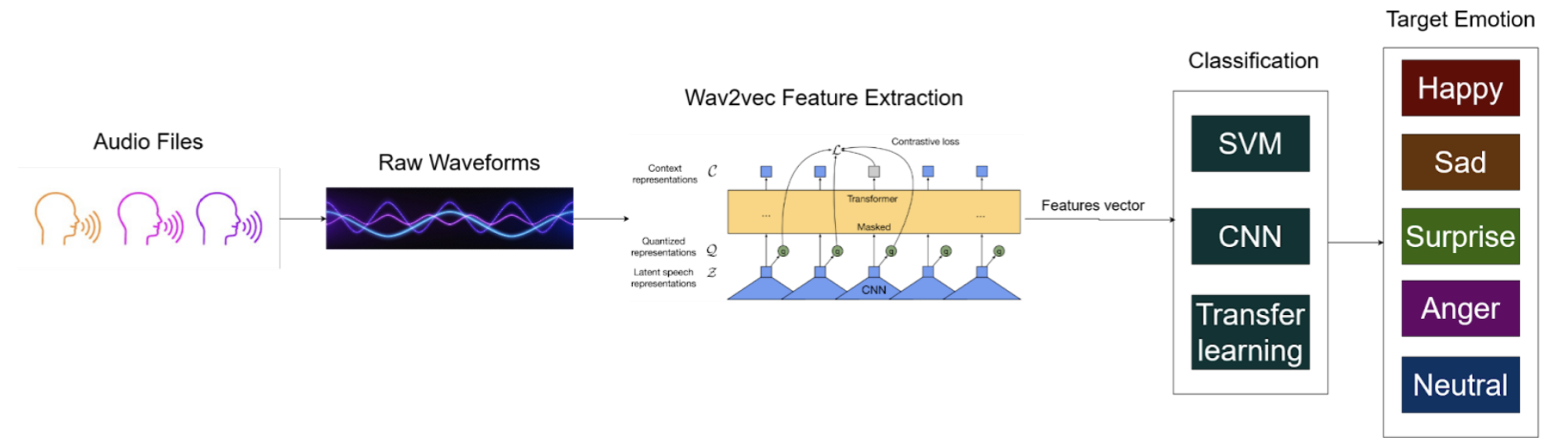}
	\caption{Structure of the model for SER paradigm}
	\label{fig1}
\end{figure*}

The three main steps of our suggested methodology are preprocessing, feature extraction, and classification. To prepare the raw audio data, the preprocessing phase first applies initial transformations. Then, a feature extractor identifies representative cues that encode emotional data. In order to predict the target emotions, classification models subsequently examine the auditory representations. Three different classifier architectures are evaluated. The data can be preprocessed before being fed into the classifiers for emotion recognition using this staged pipeline. Support vector machines (SVMs) \cite{b45}, transfer learning using prominent pre-trained models, and customized deep neural networks are all used in our categorization experiments. The pipeline is shown in \figurename \ref{fig1}. We wanted to compare performance and accuracy on the voice emotion recognition task by evaluating various classifier architectures. The main goal was to categorize the emotional state that was exhibited in each input audio clip from our dataset.

\subsection{Preprocessing}

\begin{figure}[htbp]
	\includegraphics[width=\linewidth]{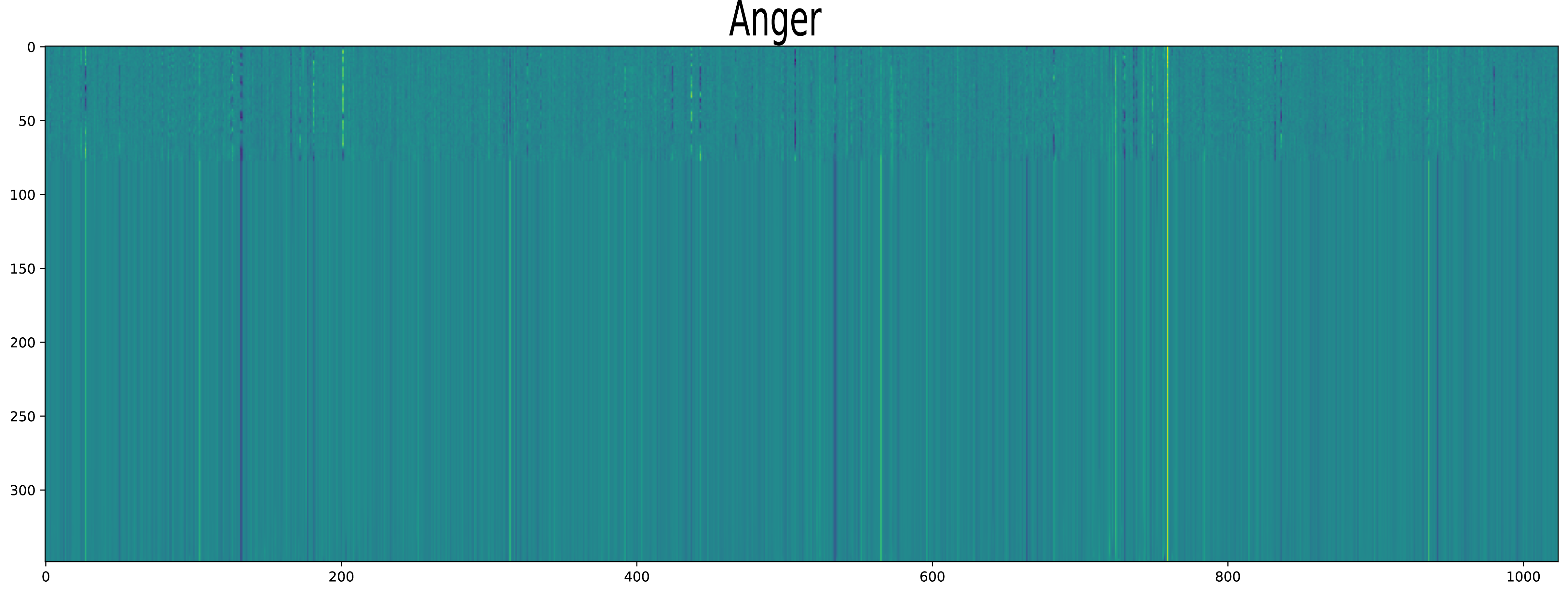}
	\caption{Output of preprocessing step}
	\label{fig2}
\end{figure}

Through zero-padding, our preprocessing standardized all audio clips to 7 seconds. The underrepresented 'fear' class was excluded. All audios are resampled to 16KHz. XLSR-53 \cite{b46}, a self-supervised multi-lingual speech representation model that has been pre-trained on 53 languages, was used to extract features. We considered the output of 24-layer feature map with 349 x 1024 feature arrays produced by XLSR-53. A sample first layer output for "sadness" is shown in \figurename \ref{fig2}. We used the mean feature vector from the first layer for the SVM classifier. We Utilized the full 349 x 1024 first layer feature map from XLSR-53 for deep neural network classifiers. CNNs received this altered data as an input in the form of a 300 x 300 2D representation. With the help of data preprocessing and the extraction of informative voice features, we improved the ability of deep networks and SVMs to classify emotions. Diverse emotional cues in the audio data were intended to be captured by the various feature representations. 

The preprocessing made it possible to use deep neural networks, which typically anticipate 2D inputs, for the task of recognizing speech emotions. Specifically, the audio data was converted into a format that was compatible with CNNs by reshaping the 349x1024 XLSR feature array into a 300x300 2D representation. These models could be used by matching the input structure to anticipated CNN architectures. Modern deep learning algorithms can be used to solve our audio emotion classification problem by first converting the audio to a matrix-shape format. The audio input was transformed by our preprocessing pipeline into a format that CNNs can use effectively.

\subsection{Classification}
Reshaping the audio features into a 300x300 matrix representation enabled leveraging deep learning techniques designed for image inputs. CNNs are well-suited for this 2D structured data, as they employ 2D convolutions to extract spatial patterns and relationships. Moreover, transfer learning can be utilized by fine-tuning models pre-trained on image datasets. Additionally, we establish baseline performance by using SVMs on the mean audio features from the first XLSR layer, as a baseline method. Although deep learning algorithms are not anticipated to match the SVM, it offers a good linear benchmark for comparison.  Hierarchical feature learning in CNN is expected to outperform the SVM model. However, the SVM offers a fundamental baseline for quantifying gains made by incorporating deep networks into our audio emotion recognition problem.

 \subsubsection{Support Vector Machines (SVM)}
 As a baseline model, we utilized a Support Vector Machine (SVM) classifier. The input representation comprised the mean feature vector from the first layer of the XLSR model, summarizing information across time similarly to the original audio We standardized the data to have zero mean, then applied a radial basis function (RBF) kernel to map the data into a higher dimensional space where a linear decision boundary could separate it. The RBF kernel allowed the SVM model to handle any nonlinear relationships between the audio features and emotion labels. By comparing deep learning approaches to this nonlinear SVM baseline, we aimed to demonstrate the benefits of learned hierarchical representations. The SVM provides a reasonable benchmark for accuracy from a non-deep model applied directly to the raw audio features.
 
 \subsubsection{Transfer Learning}
 We leveraged transfer learning as one of our deep learning approaches, utilizing EfficientNet architectures pre-trained on the ImageNet dataset. EfficientNets are a family of CNNs optimized for high accuracy with reduced parameters. We evaluated different scaled versions, including EfficientNetB0 through EfficientNetB7, which have increasing layers and capacities. The base EfficientNet model parameters were frozen, and we added new dense layers tuned for our audio emotion classification task. By reusing the pre-trained feature hierarchies in the convolutional base, we could take advantage of transferred knowledge from a source domain. Fine-tuning only the top layers adapted the model to extract emotionally relevant patterns from our preprocessed spectrograms. Compared to training a deep CNN from scratch, transfer learning provides faster convergence and better generalization with fewer training examples. Evaluating various EfficientNet scaling levels allowed us to explore trade-offs between model size, training time, and accuracy. This transfer learning approach efficiently applies deep convolution networks to recognize emotions from our reshaped audio representations.
 
 \subsubsection{Proposed CNN Model}
\begin{figure*}[htbp]
 	\includegraphics[width=\linewidth]{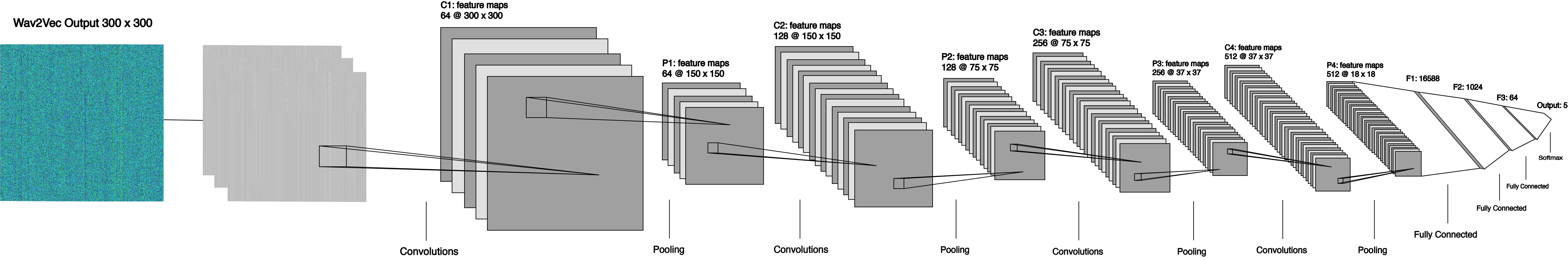}
 	\caption{Proposed CNN Model Architecture}
 	\label{fig3}
 \end{figure*}

 Finally, we propose a custom-designed CNN architecture specifically created for speech emotion classification, achieving improved performance. The model comprises seven 2D convolutional layers with progressively decreasing filter sizes, enabling the network to learn higher-level feature representations from the input spectrograms in a hierarchical manner. Batch normalization was used to increase training stability after each convolutional layer. In order to downsample the feature maps and uncover strong spatial patterns, max-pooling layers were carefully placed. Following layer pooling, dropout, and regularization were also used to reduce overfitting on the training set of data. Before a softmax classification layer, the final convolution layer outputs were flattened and routed through fully connected dense layers to compress the features shown in \tablename \ref{table1}. Our CNN model can learn a deep hierarchy of emotionally significant audio elements by increasing the filter depths and modifying the overall architecture. Our particular design seeks to achieve greater accuracy compared to traditional CNNs and transfer learning models by adjusting the network topology and hyperparameters to the specifics of our innovative preprocessed spectrogram input representation. In addition to the SVM baseline, the customized model offers a powerful deep learning strategy for generalizing across various human emotions inside the audio dataset. \figurename \ref{fig3} depicts the suggested CNN model architecture.
 
\begin{table}[]
	\caption{Proposed CNN model structure}
	\begin{tabular}{|c|c|c|}
		\hline
		layer name      & output size     & Proposed Model                                                                                                 \\ \hline
		conv1           & 300 x 300 x 64  & 3 x 3, 64, stride 1                                                                                            \\ \hline
		conv2\_x        & 150 x 150 x 128 & \begin{tabular}[c]{@{}c@{}}3 x 3 max pool, stride 1\\ {[}3 x 3, 128{]} x 2\\ {[}3 x 3, 128{]} x 2\end{tabular} \\ \hline
		conv3\_x        & 75 x 75 x 256   & \begin{tabular}[c]{@{}c@{}}{[}3 x 3, 256{]} x 2\\ {[}3 x 3, 256{]} x 2\end{tabular}                            \\ \hline
		conv4\_x        & 37 x 37 x 512   & \begin{tabular}[c]{@{}c@{}}{[}3 x 3, 512{]} x 2\\ {[}3 x 3, 512{]} x 2\end{tabular}                            \\ \hline
		conv5\_x        & 18 x 18 x 512   & {[}3 x 3, 512{]} x 2                                                                                           \\ \hline
		& 1 x 1 x 1024    & average pool                                                                                                   \\ \hline
		Fully Connected & 64              & 64 x 1024                                                                                                      \\ \hline
		softmax         & 5               &                                                                                                                \\ \hline
	\end{tabular}
	\label{table1}
\end{table}
 
\section{Experiment and Result}
\subsection{Experimental Setup}
\subsubsection{Dataset}
\begin{figure}[htbp]
	\includegraphics[width=\linewidth]{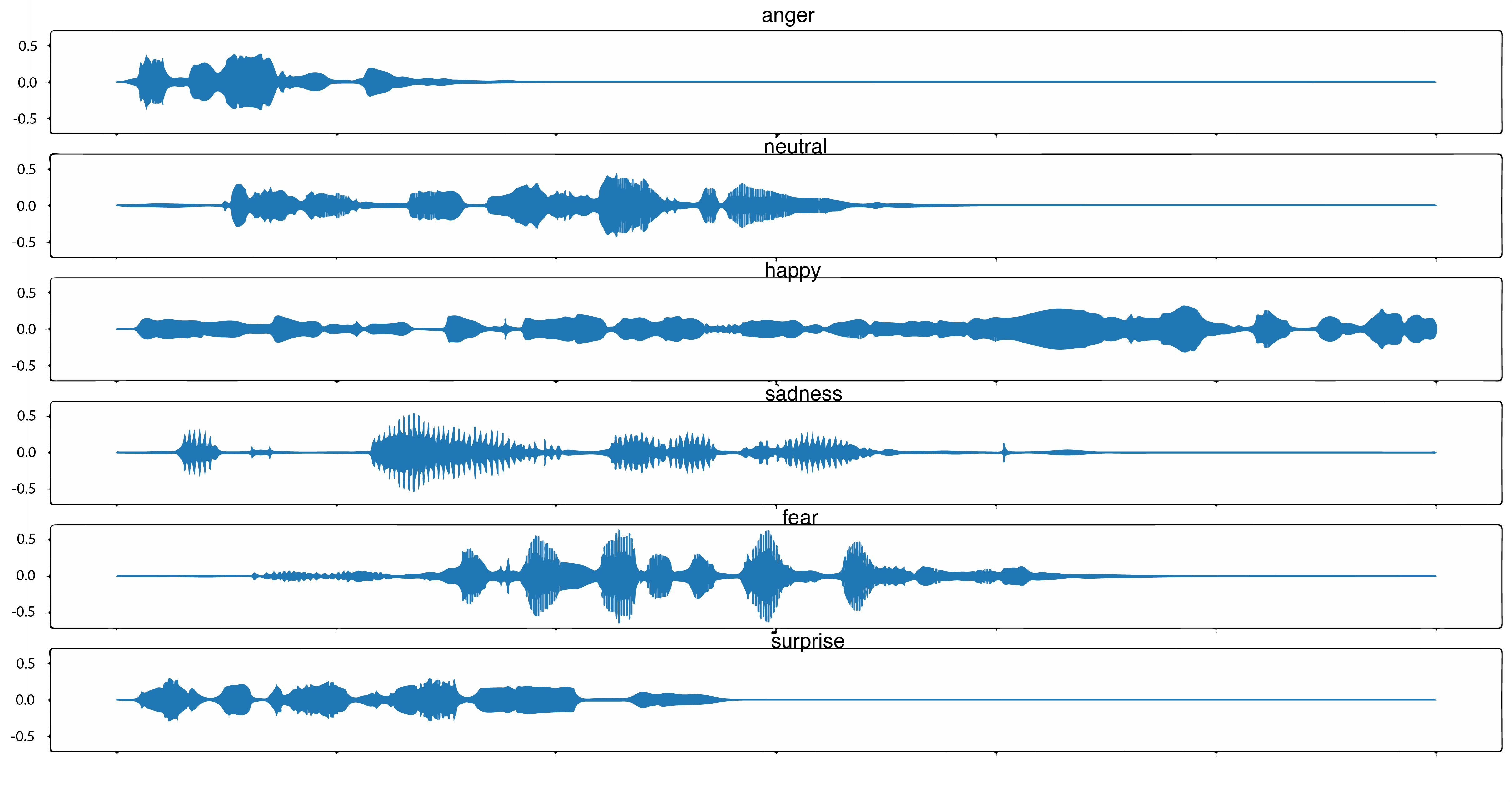}
	\caption{Random waveforms of six emotions from ShEMO dataset}
	\label{fig4}
\end{figure}

The ShEMO dataset \cite{b31} is made up of dialogues from 87 individuals that were taken from the radio and spans a duration of around 3 hours and 25 minutes. Of these speakers, 31 are women and 56 are males. This dataset has been made of 3000 utterances in the mono channel which accurance of each emotion is in \tablename \ref{table2}, and at the frequency of 44.1 kHz. We employed the following five emotions in this work: surprise, happiness, sadness, neutral, and anger. We also excluded waveforms with fear labels since there were only 38 of them (1.26 \% of the total). Additionally, audio waves longer than 7 seconds were clipped to this duration, while those shorter than 7 seconds were zero-padded to provide uniform sequence lengths, while a sample of each emotion for 3 seconds are illustrated in \figurename \ref{fig4}.

\begin{table}[h]
	\caption{Data description of ShEMO dataset}
	\begin{center}
		\resizebox{\linewidth}{!}{
			\begin{tabular}{|c|c|c|c|c|c|c|}
				\hline Emotion & Surprise & Happiness & Sadness & Anger & Neutral & Fear \\
				\hline Utterances & 225 & 201 & 449 & 1059 & 1028 & 38 \\
				\hline $\begin{array}{c}\text { Avg Length } \\
					\text { (Seconds) }\end{array}$ & 1.79 & 3.81 & 4.84 & 3.61 & 4.89 & 3.17 \\
				\hline
			\end{tabular}
		}
	\label{table2}
	\end{center}
\end{table}

\subsubsection{Parameters}
In the SVM experiment, we employed the RBF kernel \cite{b40} and the gamma was set to auto. For optimization, we employed stochastic gradient descent (SGD) \cite{b37} to train the EfficientNet-B3 network \cite{b41} and Adam \cite{b38} with a fixed learning rate of 10-3 and scheduler in our proposed model experiments. A batch size of 4 was used during training. Model evaluation was performed on full audio recordings using wav2vec representations as input, without relying on segmentation. Additionally, regularization was employed to avoid overfitting and dropout \cite{b39}—which was applied with probabilities of 0.1, 0.2, 0.3, 0.4, and 0.6—was used after the max-pooling layers. Finally, for the loss function, we used categorical cross-entropy due to the labels being tagged first.

\subsection{Results}
\begin{figure}
	\centering
	\begin{subfigure}{0.24\textwidth}
		\includegraphics[width=\textwidth]{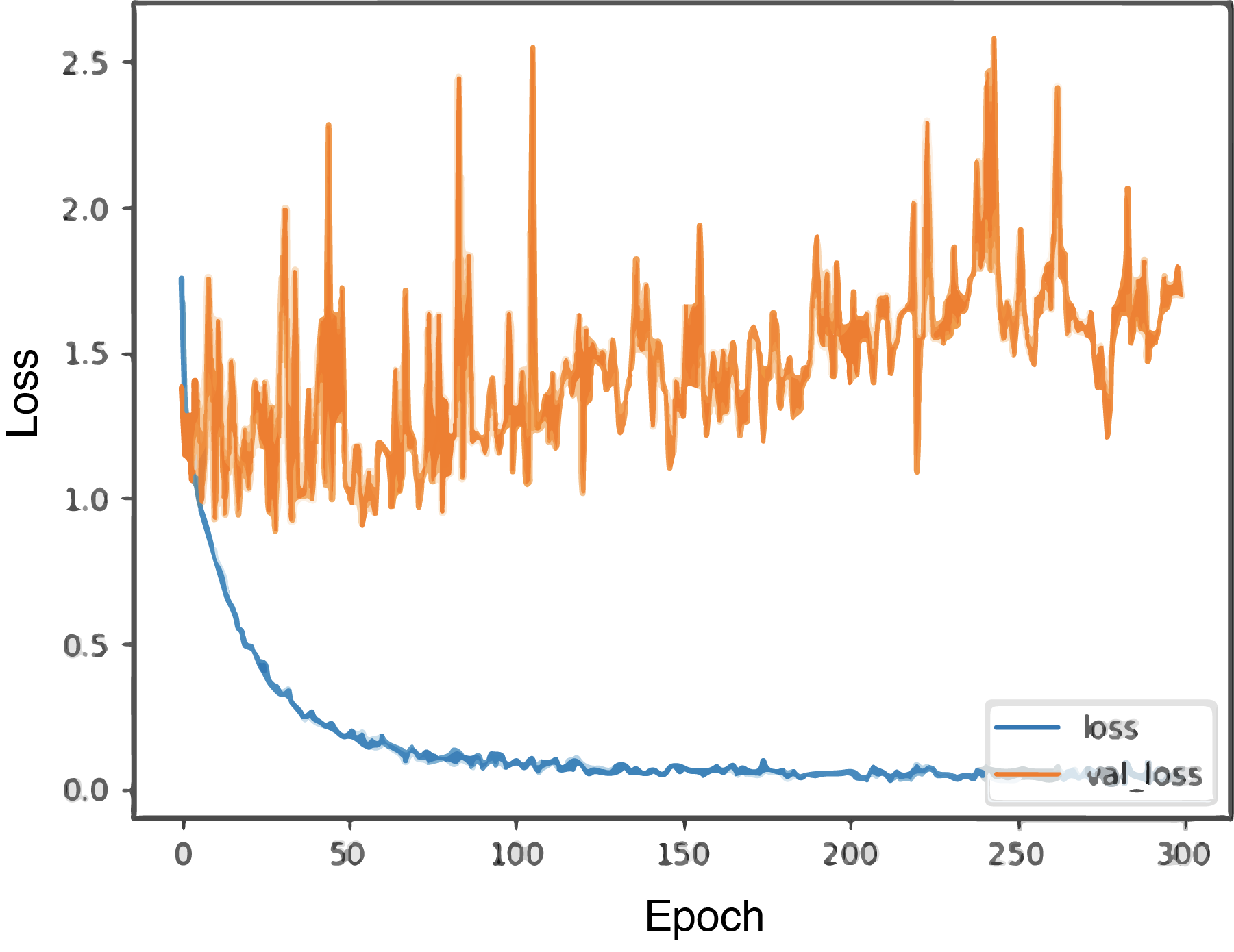}
		\label{fig:first}
	\end{subfigure}
	\hfill
	\begin{subfigure}{0.24\textwidth}
		\includegraphics[width=\textwidth]{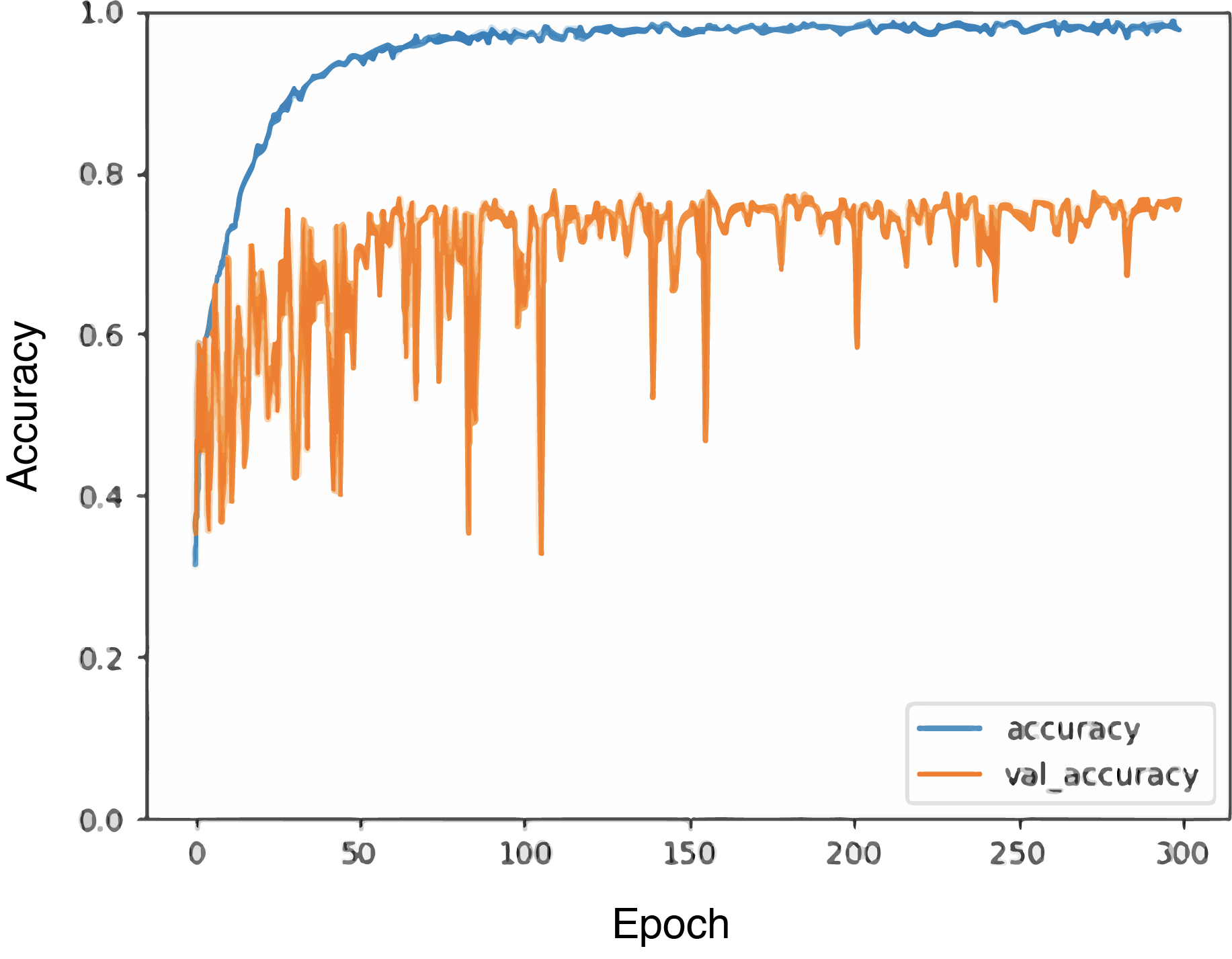}
		\label{fig:second}
	\end{subfigure}
	
	\caption{Loss and accuracy of EfficientNet model}
	\label{fig:fig5}
\end{figure}

The transfer learning loss and accuracy curves are shown in \figurename \ref{fig:fig5}. Training loss decreased as the model fit improved over epochs. However, validation loss remained nearly constant before slightly increasing after 100 epochs, indicating potential overfitting. Accuracy followed comparable trends, with training accuracy quickly reaching 77.8\% then plateauing as validation fluctuated without notable gains. The curves suggest modifications are needed to further enhance performance.

\begin{figure}
	\centering
	\begin{subfigure}{0.24\textwidth}
		\includegraphics[width=\textwidth]{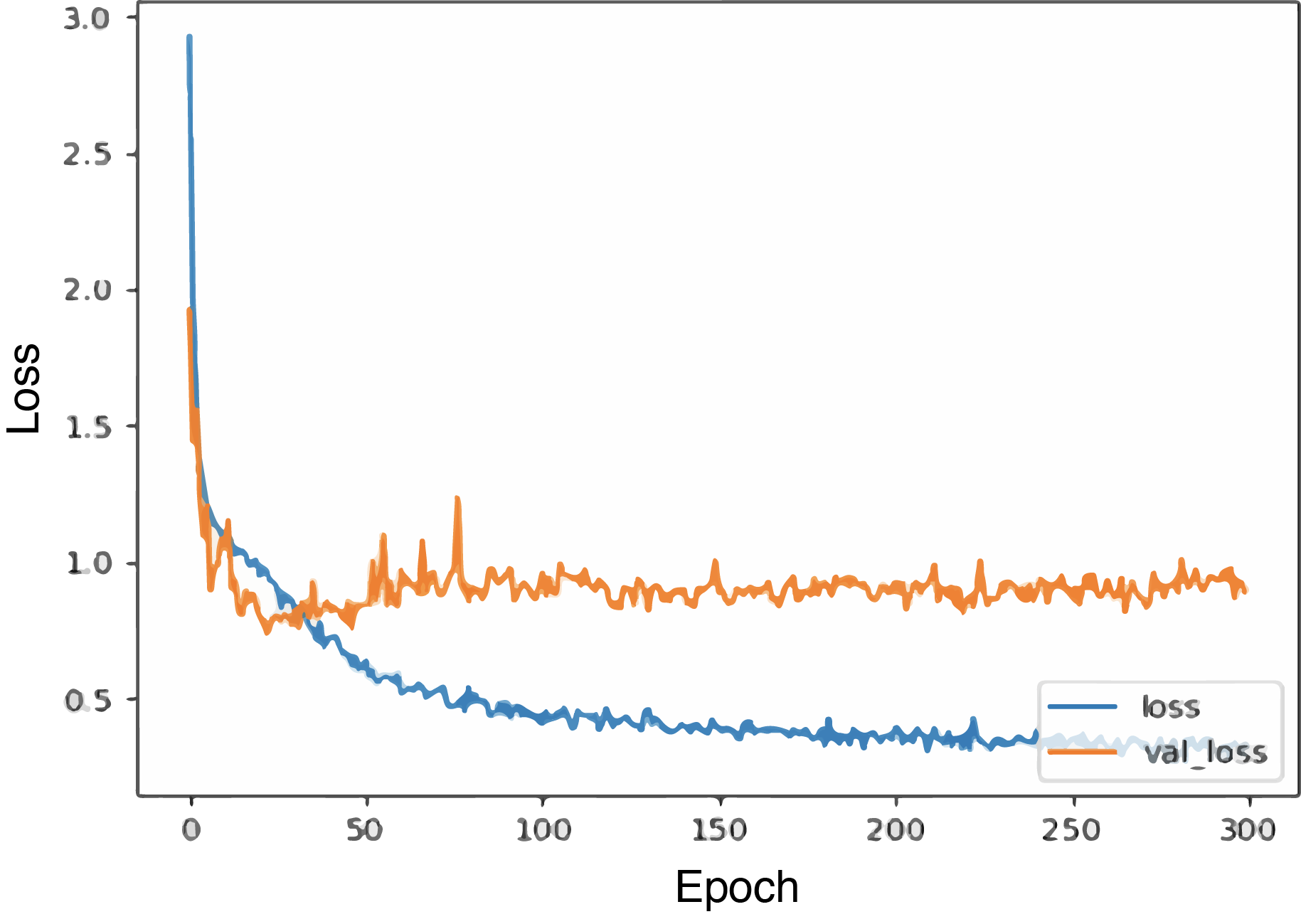}
		\label{fig:first2}
	\end{subfigure}
	\hfill
	\begin{subfigure}{0.24\textwidth}
		\includegraphics[width=\textwidth]{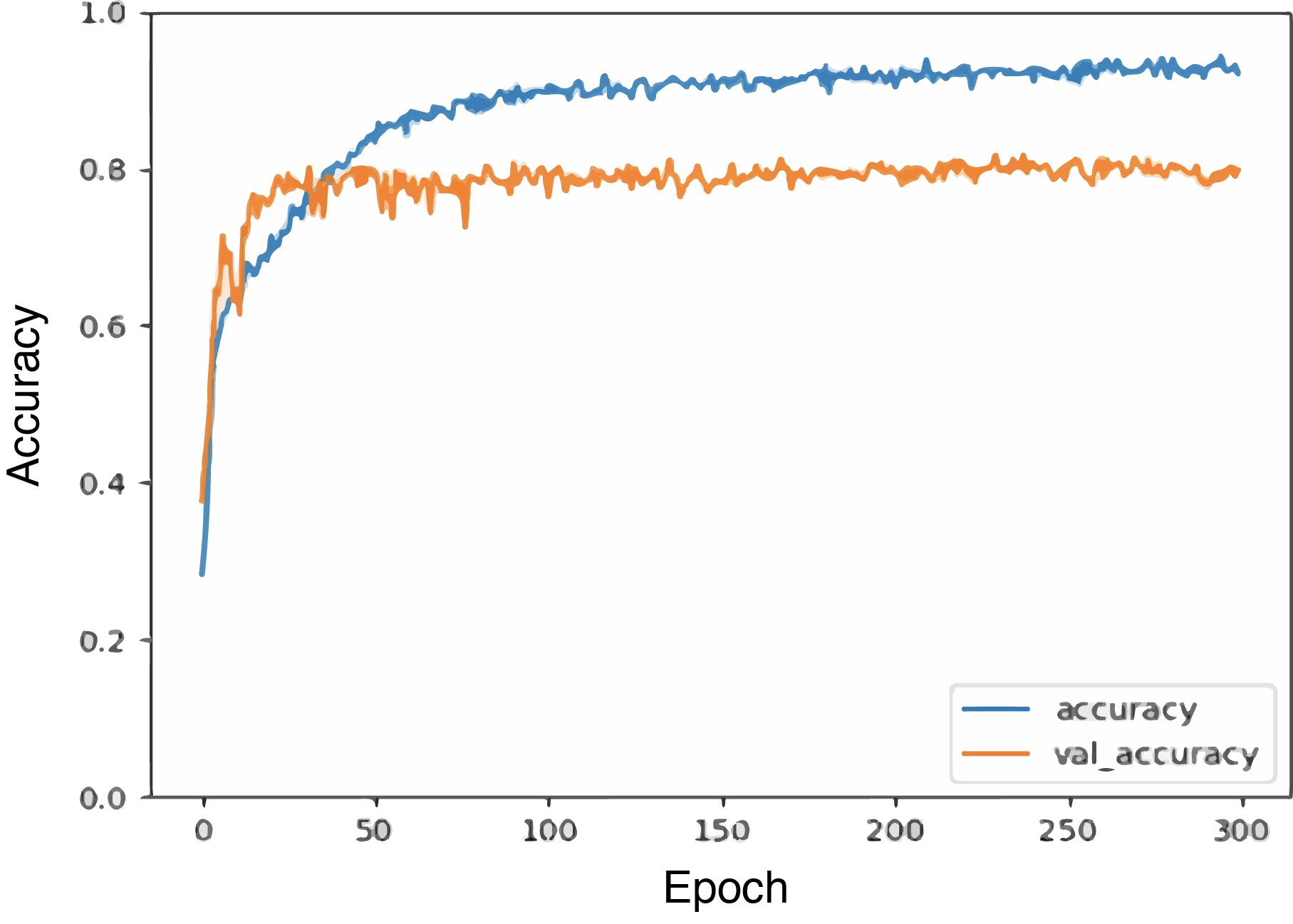}
		\label{fig:second2}
	\end{subfigure}
	
	\caption{Loss and accuracy of proposed model}
	\label{fig:fig6}
\end{figure}

As shown in \figurename \ref{fig:fig6}, the CNN model exhibited decreasing training and validation loss over epochs, indicating successful learning from the data. The validation loss remained lower than training loss throughout training, suggesting effective generalization to new data. Both training and validation accuracy steadily increased over epochs. In later epochs, validation accuracy surpassed 80\%, outperforming SVM and transfer learning models. These curves demonstrate our proposed architecture achieved higher precision and lower loss on training and validation sets compared to prior benchmarks. Moreover, superior validation over training performance highlights model generalizability. 

We compare our results to prior work by Keesing et al. \cite{b42}, which utilized wav2vec features on the ShEMO dataset. Our hypotheses are that expanding model capacity and depth can improve performance. As shown in \tablename \ref{table3}, our proposed model achieves superior accuracy on the ShEMO validation and test sets compared to \cite{b42}. Specifically, our test set recall improves by an absolute value of approximately 13\%.

\vspace{-3 mm}
\begin{table}[h]
	\caption{Performance comparison between the proposed method and other state-of-the-art methods}
	\begin{center}
		\resizebox{\linewidth}{!}{
			\begin{tabular}{|l|l|l|l|}
				\hline & Accuracy & F1 score & Recall \\
				\hline Keesing et al. \cite{b42} & - & - & 0.640 \\
				\hline Yazdani et al. \cite{b11} & $78.29 \%$ & - & - \\
				\hline Sharma et al. \cite{b44} & - & 0.860 & 0.862 \\
				\hline SVM & $76 \%$ & 0.73 & 0.76 \\
				\hline Efficientnet-B3 & $78.7 \%$ & 0.75 & 0.77 \\
				\hline Proposed & $81.7 \%$ & 0.87 & 0.870 \\
				\hline
			\end{tabular}
		}
		\label{table3}
	\end{center}
	
\end{table}
We also compare against prior work utilizing hand-crafted features on ShEMO, summarized in Table 2. Notably, Yazdani et al. \cite{b44} achieved state-of-the-art accuracy using this approach. Our model surpasses their accuracy by 3.41\%. Critically, unlike Yazdani et al.'s reliance on manually engineered features like MFCCs, our model utilizes an end-to-end architecture, operating directly on raw waveform inputs independent of temporal or spectral characteristics.

Sharma et al. \cite{b44} presents the most recent benchmark on the full ShEMO dataset, achieving state-of-the-art results. Their reported test set weighted average recall is 0.862, just 0.8\% absolutely lower than our model utilizing wav2vec features. For completeness, their F1 score is 0.860 compared to 0.870 for our proposed method, a 0.01 absolute improvement. By incorporating wav2vec representations into a deeper CNN, our approach advances the state-of-the-art in this dimension.

\section{Conclusion}
In this study, we introduced a novel approach to SER by combining self-supervised feature extraction with a CNN architecture. Firstly, by harnessing wav2vec feature extraction and a costume designed CNN-based model, we achieved substantial improvements over prior SER methods. Notably, our proposed approach outperformed the SVM-based model and transfer learning methods, showcasing superior generalization capabilities. Comparing our model to prior work, we observed substantial performance enhancements. In particular, we surpassed the accuracy achieved by models relying on hand-crafted features, such as MFCCs, underscoring the potential of end-to-end architectures that operate directly on raw waveform inputs. Furthermore, our approach demonstrated competitive results when benchmarked against the most recent state-of-the-art methods. Our findings contribute to the evolving landscape of SER, emphasizing the importance of data-driven approaches in advancing the understanding of emotional cues in speech.


\begin{thebibliography}{00}
\bibitem{b1} Arora, S. J. and R. P. Singh (2012). "Automatic speech recognition: a review." International Journal of Computer Applications 60(9).
\bibitem{b2} Ververidis, D. and C. Kotropoulos (2006). "Emotional speech recognition: Resources, features, and methods." Speech communication 48(9): 1162-1181.
\bibitem{b3} Sánchez-Gutiérrez, M. E., et al. (2014). Deep learning for emotional speech recognition. Pattern Recognition: 6th Mexican Conference, MCPR 2014, Cancun, Mexico, June 25-28, 2014. Proceedings 6, Springer.
\bibitem{b4} Shashidhar, G., K. Koolagudi and R. Sreenivasa (2012). "Emotion recognition from speech: a review." Springer Science+ Business Media 15: 99-117.
\bibitem{b5} El Ayadi, M., M. S. Kamel and F. Karray (2011). "Survey on speech emotion recognition: Features, classification schemes, and databases." Pattern recognition 44(3): 572-587.
\bibitem{b6} Khalil, R. A., et al. (2019). "Speech emotion recognition using deep learning techniques: A review." IEEE Access 7: 117327-117345.
\bibitem{b7} Deng, L. and D. Yu (2014). "Deep learning: methods and applications." Foundations and trends in signal processing 7(3–4): 197-387.
\bibitem{b8} Schuller, B., G. Rigoll and M. Lang (2004). Speech emotion recognition combining acoustic features and linguistic information in a hybrid support vector machine-belief network architecture. 2004 IEEE international conference on acoustics, speech, and signal processing, IEEE.
\bibitem{b9} France, D. J., et al. (2000). "Acoustical properties of speech as indicators of depression and suicidal risk." IEEE transactions on Biomedical Engineering 47(7): 829-837.
\bibitem{b10}Jiang, W., et al. (2019). "Speech emotion recognition with heterogeneous feature unification of deep neural network." Sensors 19(12): 2730.
\bibitem{b11} Yazdani, A., H. Simchi and Y. Shekofteh (2021). Emotion recognition in persian speech using deep neural networks. 2021 11th International Conference on Computer Engineering and Knowledge (ICCKE), IEEE.
\bibitem{b12} Burget, L., Schwarz, P., Agarwal, M., Akyazi, P., Feng, K., Ghoshal, A., ... \& Thomas, S. (2010, March). Multilingual acoustic modeling for speech recognition based on subspace Gaussian mixture models. In 2010 IEEE international conference on acoustics, speech and signal processing (pp. 4334-4337). IEEE.
\bibitem{b13} Conneau, A., Baevski, A., Collobert, R., Mohamed, A., \& Auli, M. (2020). Unsupervised cross-lingual representation learning for speech recognition. arXiv preprint arXiv:2006.13979.
\bibitem{b14} Baevski, A., Zhou, Y., Mohamed, A., \& Auli, M. (2020). wav2vec 2.0: A framework for self-supervised learning of speech representations. Advances in neural information processing systems, 33, 12449-12460.
\bibitem{b15} Lee, J. and I. Tashev (2015). High-level feature representation using recurrent neural network for speech emotion recognition. Interspeech 2015.
\bibitem{b16} M. P. Lewis, G. F. Simon, and C. D. Fennig. Ethnologue: Languages of the world, nineteenth edition. Online version: http://www.ethnologue.com, 2016.
\bibitem{b17} Zhang, Y., Park, D. S., Han, W., Qin, J., Gulati, A., Shor, J., ... \& Wu, Y. (2022). Bigssl: Exploring the frontier of large-scale semi-supervised learning for automatic speech recognition. IEEE Journal of Selected Topics in Signal Processing, 16(6), 1519-1532.
\bibitem{b18} Banse, R. and K. R. Scherer (1996). "Acoustic profiles in vocal emotion expression." Journal of personality and social psychology 70(3): 614.
\bibitem{b19} Pepino, L., Riera, P., \& Ferrer, L. (2021). Emotion recognition from speech using wav2vec 2.0 embeddings. arXiv preprint arXiv:2104.03502.
\bibitem{b20} Brave, S., \& Nass, C. (2007). Emotion in human-computer interaction. In The human-computer interaction handbook (pp. 103-118). CRC Press.
\bibitem{b21} Li, X., Dalmia, S., Li, J., Lee, M., Littell, P., Yao, J., ... \& Metze, F. (2020, May). Universal phone recognition with a multilingual allophone system. In ICASSP 2020-2020 IEEE International Conference on Acoustics, Speech and Signal Processing (ICASSP) (pp. 8249-8253). IEEE.
\bibitem{b22} McFee, B., Raffel, C., Liang, D., Ellis, D. P., McVicar, M., Battenberg, E., \& Nieto, O. (2015, July). librosa: Audio and music signal analysis in python. In Proceedings of the 14th python in science conference (Vol. 8, pp. 18-25).
\bibitem{b23} Tzirakis, P., Zhang, J., \& Schuller, B. W. (2018, April). End-to-end speech emotion recognition using deep neural networks. In 2018 IEEE international conference on acoustics, speech and signal processing (ICASSP) (pp. 5089-5093). IEEE.
\bibitem{b24} Wöllmer, M., Eyben, F., Reiter, S., Schuller, B., Cox, C., Douglas-Cowie, E., \& Cowie, R. (2008). Abandoning emotion classes-towards continuous emotion recognition with modelling of long-range dependencies.
\bibitem{b25} Stuhlsatz, A., Meyer, C., Eyben, F., Zielke, T., Meier, G.,\& Schuller, B. (2011, May). Deep neural networks for acoustic emotion recognition: Raising the benchmarks. In 2011 IEEE international conference on acoustics, speech and signal processing (ICASSP) (pp. 5688-5691). IEEE.
\bibitem{b26} Ververidis, D., Kotropoulos, C., \& Pitas, I. (2004, May). Automatic emotional speech classification. In 2004 IEEE international conference on acoustics, speech, and signal processing (Vol. 1, pp. I-593). IEEE.
\bibitem{b27} Xiao, Z., Dellandrea, E., Dou, W., \& Chen, L. (2005, September). Features extraction and selection for emotional speech classification. In IEEE Conference on Advanced Video and Signal Based Surveillance, 2005. (pp. 411-416). IEEE.
\bibitem{b28} Pan, Y., Shen, P., \& Shen, L. (2005, September). Feature extraction and selection in speech emotion recognition. In IEEE Conference on Advanced Video and Signal Based Surveillance (AVSS 2005), Como, Italy.
\bibitem{b29} Zulkifly, M. A. A., \& Yahya, N. (2017, September). Relative spectral-perceptual linear prediction (RASTA-PLP) speech signals analysis using singular value decomposition (SVD). In 2017 IEEE 3rd International Symposium in Robotics and Manufacturing Automation (ROMA) (pp. 1-5). IEEE.
\bibitem{b30} Zhang, Y., et al. (2018). Attention based fully convolutional network for speech emotion recognition. 2018 Asia-Pacific Signal and Information Processing Association Annual Summit and Conference (APSIPA ASC), IEEE.
\bibitem{b31} Mohamad Nezami, O., P. Jamshid Lou and M. Karami (2019). "ShEMO: a large-scale validated database for Persian speech emotion detection." Language Resources and Evaluation 53: 1-16.
\bibitem{b32} O'Shea, K. and R. Nash (2015). "An introduction to convolutional neural networks." arXiv preprint arXiv:1511.08458.
\bibitem{b33} Elgeish, “https://huggingface.co/elgeish/wav2vec2-large-xlsr-53-arabic,” 2020.
\bibitem{b34} Peter, C., \& Urban, B. (2012). Emotion in human-computer interaction. Expanding the frontiers of visual analytics and visualization, 239-262.	
\bibitem{b35} Myers, B. A. (1998). A brief history of human-computer interaction technology. interactions, 5(2), 44-54.
\bibitem{b36} Babu, A., Wang, C., Tjandra, A., Lakhotia, K., Xu, Q., Goyal, N., ... \& Auli, M. (2021). XLS-R: Self-supervised cross-lingual speech representation learning at scale. arXiv preprint arXiv:2111.09296.	
\bibitem{b37} Ruder, S. (2016). An overview of gradient descent optimization algorithms. arXiv preprint arXiv:1609.04747.
\bibitem{b38} Kingma, D. P., \& Ba, J. (2014). Adam: A method for stochastic optimization. arXiv preprint arXiv:1412.6980.
\bibitem{b39} Srivastava, N., Hinton, G., Krizhevsky, A., Sutskever, I., \& Salakhutdinov, R. (2014). Dropout: a simple way to prevent neural networks from overfitting. The journal of machine learning research, 15(1), 1929-1958.
\bibitem{b40} Patle, A., \& Chouhan, D. S. (2013, January). SVM kernel functions for classification. In 2013 International conference on advances in technology and engineering (ICATE) (pp. 1-9). IEEE.
\bibitem{b41} Tan, M., \& Le, Q. (2019, May). Efficientnet: Rethinking model scaling for convolutional neural networks. In International conference on machine learning (pp. 6105-6114). PMLR.
\bibitem{b42} Keesing, A., Koh, Y. S., \& Witbrock, M. (2021, August). Acoustic Features and Neural Representations for Categorical Emotion Recognition from Speech. In Interspeech (pp. 3415-3419).
\bibitem{b43} M. G. de Pinto, M. Polignano, P. Lops, and G. Semeraro, “Emotions Understanding Model from Spoken Language using Deep Neural Networks and Mel-Frequency Cepstral Coefficients,” in 2020 IEEE Conference on Evolving and Adaptive Intelligent Systems (EAIS), May 2020, pp. 1–5
\bibitem{b44} Sharma, M. (2022). Multi-lingual multi-task speech emotion recognition using wav2vec 2.0. ICASSP 2022-2022 IEEE International Conference on Acoustics, Speech and Signal Processing (ICASSP), IEEE.
\bibitem{b45} Cortes, C., \& Vapnik, V. (1995). Support-vector networks. Machine learning, 20, 273-297.

\bibitem{b46} Babu, A., Wang, C., Tjandra, A., Lakhotia, K., Xu, Q., Goyal, N., ... \& Auli, M. (2021). XLS-R: Self-supervised cross-lingual speech representation learning at scale. arXiv preprint arXiv:2111.09296.

\end{thebibliography}
\end{document}